\documentclass{cimento}

\def\beq{\begin{equation}}
\def\eeq{\end{equation}}
\def\bea{\begin{eqnarray}}
\def\eea{\end{eqnarray}}

\def\nn{\nonumber}
\def\Eq#1{Eq.~(\ref{#1})}

\usepackage{graphicx}  

\title{Charge asymmetry: a theory appraisal}
\author{G.~Rodrigo\from{ins:x}\thanks{E-mail: german.rodrigo@ific.uv.es},
\atque P.~Ferrario\from{ins:x}\thanks{E-mail: paola.ferrario@ific.uv.es}}
\instlist{\inst{ins:x} Instituto de F\'{\i}sica Corpuscular, UVEG --
Consejo Superior de Investigaciones Cient\'{\i}ficas, \\ 
Parc Cient\'{\i}fic de Paterna, 
Apartado de Correos 22085, E-46071 Valencia, Spain.}
\PACSes{
\PACSit{14.65.Ha}{Top quarks}
\PACSit{11.30.Er}{Charge conjugation, parity, time reversal, and other
  discrete symmetries}
\PACSit{12.10.Dm}{Unified theories and models of strong and electroweak 
interactions}}

\begin{document}

\maketitle

\begin{abstract}
The most recent measurements at Tevatron of the charge asymmetry 
in top-antitop quark pair production reduce the discrepancy 
with the Standard Model from 2$\sigma$ to 1.7$\sigma$, and open 
a little window,  at 95\% C.L., for negative contributions 
to the charge asymmetry beyond the SM. 
We update our analysis for colour
octet gauge bosons or axigluons in flavour universal and 
flavour non universal scenarios. We review other possible 
models and make an educated guess on their parameter
space allowed by the new measurements.  
Finally, we comment on the prospects to measure the charge 
asymmetry at the LHC.
\end{abstract}

\section{Introduction}

The top quark, being the heaviest known elementary
particle, plays a fundamental role in many extensions of
the Standard Model (SM) and in alternative mechanisms
for the electroweak symmetry breaking (EWSB). Since its
discovery in 1995 at Tevatron, many properties of the
top quark, such as mass and total cross-section, have
been measured with high precision, allowing also to set 
limits on physics beyond the SM.

The LHC plans to collect $1$~fb$^{-1}$ of data 
at $7$~TeV centre of mass energy by the end of 2011. 
At that energy the total cross-section for top-antitop 
quark pair production is about $160$~pb~\cite{Ahrens:2010zv}; 
thus a sample of about $10^5$ top quark pairs will be available 
by the end of 2011 to perform high precision measurements, 
besides offering new opportunities to probe new physics in 
the top quark sector.
Moreover, a significant fraction of top-antitop quark events will be 
produced in association with jets. 

Several models predict the existence of new electroweak 
$W'$ and $Z'$ gauge bosons, colour-octet gauge bosons, 
coloured scalars or gravitons that should
be detectable in top-antitop quark events, particularly in those
models where the coupling of the new states to the third
generation is enhanced with respect to the lighter fermions.
Direct searches at Tevatron~\cite{masslimits} set lower bounds on 
the mass of colorons and flavour universal axigluons 
at about 1.2~TeV, at about 700 to 800~GeV for extra weak 
boson, and at about 500~GeV for gravitons. 
An interesting and powerful observable to distinguish among 
different models is the charge asymmetry. 

\begin{figure}
\begin{center}
\includegraphics[width=13cm]{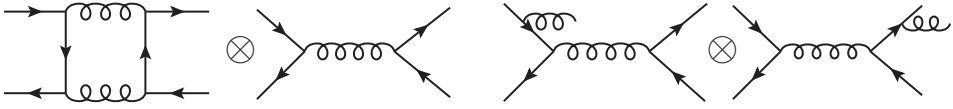}
\caption{Origin of the QCD charge asymmetry.\label{fig:aFBasym} }
\end{center}
\end{figure}

\section{Charge asymmetry in QCD}

At leading order in QCD the differential distributions of top and
antitop quarks are identical. But due to higher order radiative corrections
(Fig.~\ref{fig:aFBasym}) a charge asymmetry is generated
at ${\cal O} (\alpha_S^3)$ in $q\bar q$ events, and top quarks
become more abundant in the direction of the incoming light quarks.
At Tevatron, the charge asymmetry is equivalent to a forward--backward 
asymmetry, due to the charge conjugation symmetry. 
Chromoelectric and chromomagnetic contributions do not generate 
any asymmetry. The QCD prediction for Tevatron,
including a small mixed QCD-electroweak contribution,
 is~\cite{Antunano:2007da,Bowen:2005ap,mynlo}
\beq
A^{p\bar p} = \frac{N_t(y\ge 0)-N_{\bar t}(y\ge 0)}
{N_t(y\ge 0)+N_{\bar t}(y\ge 0)} = 0.051(6)~,
\label{appbar}
\eeq
where $y$ denotes the rapidity. 
The charge asymmetry can also be defined through
$\Delta y=y_t-y_{\bar t}$, which is equivalent to evaluate
the asymmetry in the $t\bar t$ rest frame because $\Delta y$
is invariant under boosts. In that frame the asymmetry
is about $50$\% larger~\cite{Antunano:2007da}:
$A^{t\bar t} = 0.078(9)$.
Recent threshold resummations~\cite{Ahrens:2010zv,Almeida:2008ug}
shift the central values for the inclusive asymmetries
by a few per mille only.

The most recent measurements from CDF~\cite{ICHEPCDF}, 
with $5.3$~fb$^{-1}$, are in both frames
\bea
&& A^{p\bar p} = 0.150 \pm 0.050_{\rm stat.} \pm 0.024_{\rm syst.}~, \nn \\
&& A^{t\bar t} = 0.158 \pm 0.072_{\rm stat.} \pm 0.017_{\rm syst.}~,
\label{renewcdf}
\eea
respectively. The measurement presented by D0~\cite{ICHEPD0},
with $4.3$~fb$^{-1}$ and in the observed region, is
\bea
&& A^{t\bar t}_{\rm obs.} = 0.08 \pm 0.04_{\rm stat.} \pm 0.01_{\rm syst.}~.
\label{renewd0}
\eea
With respect to the previously published 
results~\cite{Aaltonen:2008hc,:2007qb},
the new measurements are more in agreement with the SM. 
If we take the CDF result as reference, the discrepancy with respect 
to the SM has been reduced from 2$\sigma$ to 1.7$\sigma$. Moreover, 
while vanishing or negative contributions to the asymmetry 
were disfavoured at 95\% C.L. previously, the new measurements
open a little window for negative asymmetries beyond the SM. 

\section{Colour-octet gauge bosons}

\begin{figure}
\begin{center}
\includegraphics[width=7cm]{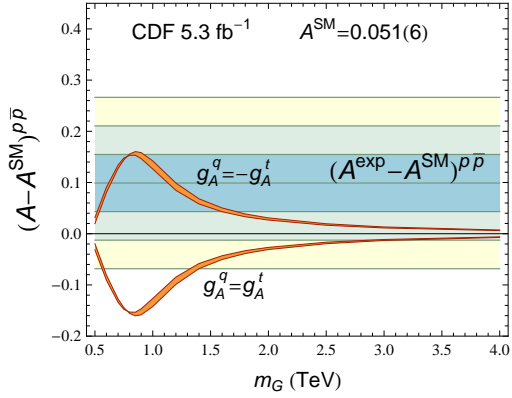} 
\caption{\label{fig:anti} Comparison of the axigluon contribution 
($g_V^q=g_V^t=0, g_A^q=g_A^t=1$) to the top quark charge asymmetry 
with the $1\sigma$, $2\sigma$ and $3\sigma$
contours as a function of the axigluon mass. 
The case $g_A^q=-g_A^t=1$ is also shown.}
\end{center}
\end{figure}

Colour-octet gauge bosons appear in chiral colour models~\cite{chiralcolor},
where the SM colour group have been extended 
to $SU(3)_R \otimes SU(3)_L$, and the symmetry
breaking to the diagonal $SU(3)_C$ generates the massive axigluon,
which couples to quarks with a pure axial-vector structure and
the same strength as QCD.
Chiral colour models require also the existence of extra fermions
to cancel anomalies, and extra Higgs bosons to break the enlarged
gauge symmetry. The extra states, however, are usually assumed to be  
arbitrary heavy. Those models can also be generalised
by considering different coupling constants associated with each
$SU(3)$ component~\cite{Cuypers:1990hb,Carone:2008rx,
Martynov:2009en,Zerwekh:2009vi},
thus generating both vector and axial-vector couplings
of the axigluon to quarks. If the two copies of the 
$SU(3)$ group are non chiral, the new gauge boson is known as 
coloron and couple only vectorially to quarks~\cite{colorons,Choudhury:2007ux}.
Massive gluons also appear as Kaluza-Klein~\cite{KK} excitations 
in models of extra dimensions~\cite{Randall:1999ee}.

In the most general scenario 
a colour-octet resonance $G_\mu^a$ interacts with quarks 
with arbitrary vector $g_V^{q_i}$ and axial-vector $g_A^{q_i}$ strength  
relative to the strong coupling $g_S$: 
\beq
{\cal L} = g_S \, t^a \, 
\bar{q}_i (g_V^{q_i} + g_A^{q_i} \, \gamma_5) \, 
\gamma^\mu \, G_\mu^a \, q_i~.
\label{eq:G}
\eeq
In explicit models, parity, gauge invariance or orthonormality 
of field profiles prevent a direct coupling of $G_\mu^a$
to an even number gluons; thus it is natural to assume 
that the extra gauge boson do not modify gluon-gluon fusion. 

The Born cross-section for $q\bar{q}$ annihilation into top quarks 
in the presence of a colour-octet vector resonance 
reads~\cite{Ferrario:2008wm}
\bea
\frac{d\sigma^{q\bar{q}\rightarrow t \bar{t}}}{d\cos \hat{\theta}} &=&
\alpha_S^2 \: \frac{T_F C_F}{N_C} \:
\frac{\pi \beta}{2 \hat{s}} \bigg\{ 1+c^2+4m^2
+ \frac{2 \hat{s} (\hat{s}-m_G^2)}
{(\hat{s}-m_G^2)^2+m_G^2 \Gamma_G^2}
\left[ g_V^q \, g_V^t \, (1+c^2+4m^2) \right. \nn \\ && \left. 
+ 2 \, g_A^q \, g_A^t \, c  \right] +
\frac{\hat{s}^2} {(\hat{s}-m_G^2)^2+m_G^2 \Gamma_G^2}
\left[ \left( (g_V^q)^2+(g_A^q)^2 \right) \right. \nn \\ && \times 
\left( (g_V^t)^2 (1+c^2+4m^2) +  (g_A^t)^2 (1+c^2-4m^2) \right)
\left.\left.
+ 8 \, g_V^q \, g_A^q \, g_V^t \, g_A^t \, c \, \right]
\right\}~,
\label{eq:bornqq}
\eea
where $\hat{\theta}$ is the polar angle of the top quark with respect
to the incoming quark in the centre of mass rest frame,
$\hat{s}$ is the squared partonic invariant mass,
$T_F=1/2$, $N_C=3$ and $C_F=4/3$ are colour factors,
$\beta = \sqrt{1-4m^2}$ is the velocity of the top quark,
with $m=m_t/\sqrt{\hat{s}}$, and $c = \beta \cos \hat{\theta}$.
The parameters $g_V^q (g_V^t)$ and $g_A^q(g_A^t)$ represent, 
respectively, the vector and axial-vector couplings of the
excited gluons to the light quarks (top quarks).
Colour-octet vector resonances are naturally broad: 
$\Gamma_G/m_G = {\cal O} (\alpha_S)$~. 

The terms in \Eq{eq:bornqq} that are odd in $c$ generate the 
charge asymmetry. Due to the factor $(\hat{s}-m_G^2)$ the charge 
asymmetry generated in flavour universal models, $g_A^q=g_A^t$,  
is in general negative. A positive asymmetry can be generated 
if $g_A^q g_A^t <0$~\cite{Ferrario:2009bz,phd,Frampton:2009ve}, 
or if the term $8 g_V^q g_A^t g_V^q g_A^t c$ dominates over 
the interference. 

\begin{figure}
\begin{center}
\includegraphics[width=6cm]{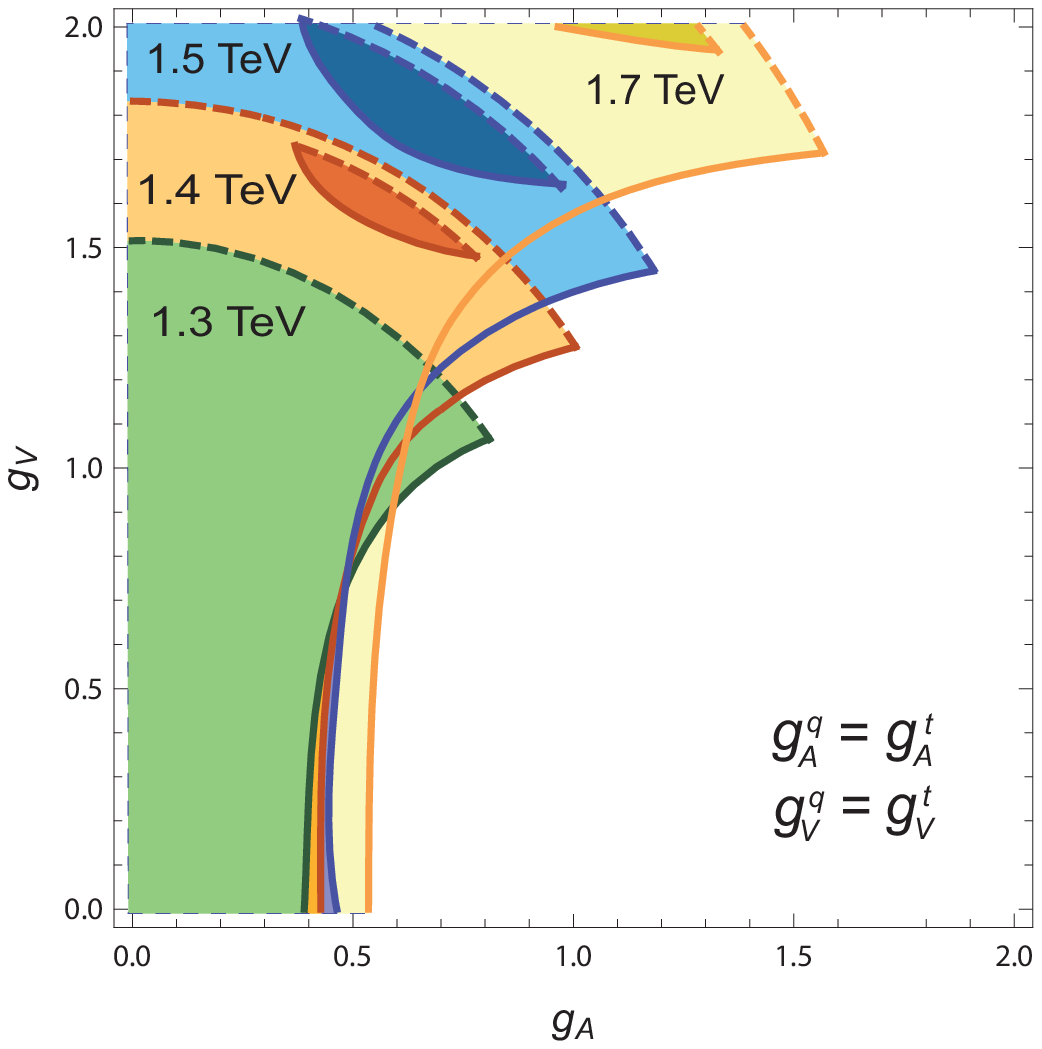} 
\includegraphics[width=6cm]{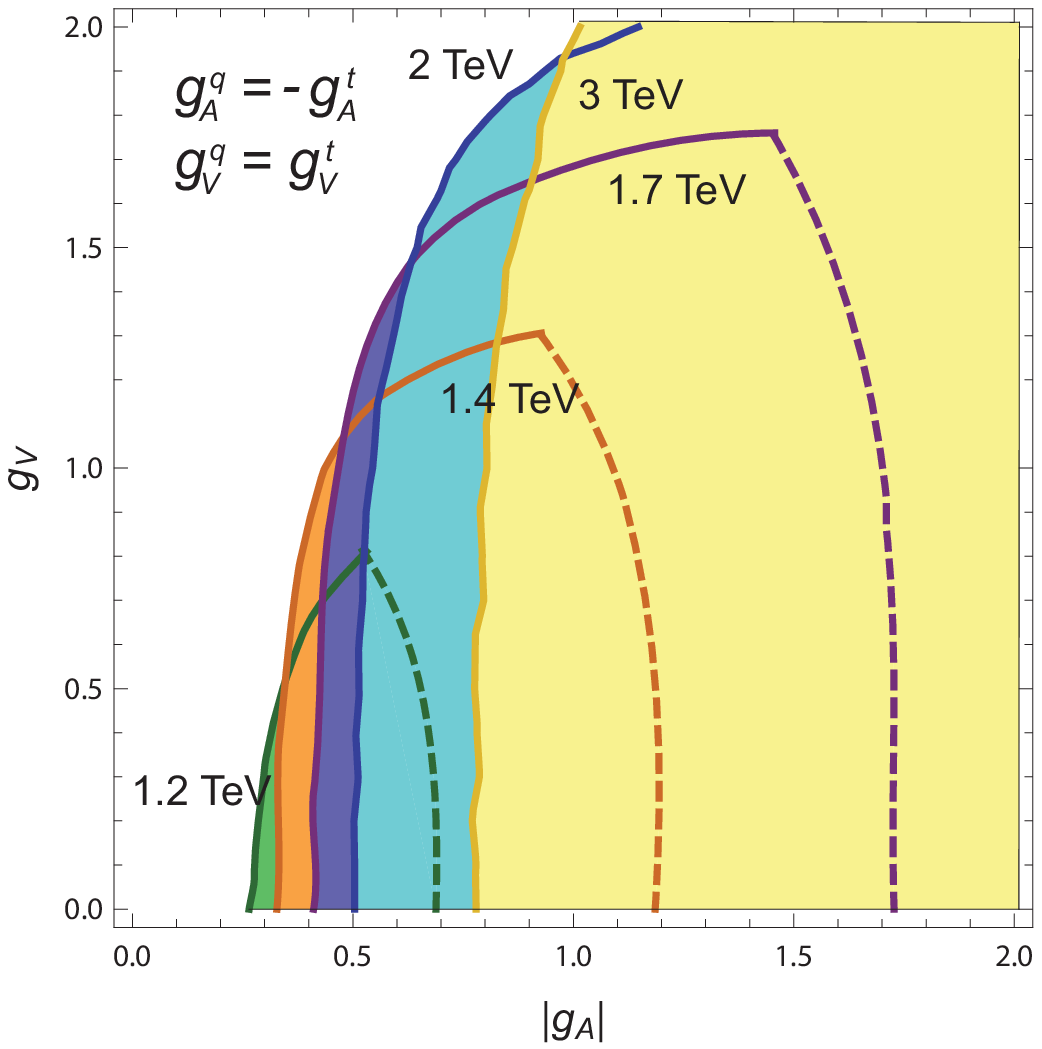} 
\caption{\label{fig:universal} Contours at 90\% and 95\% C.L. as
a function of the vector and axial-vector couplings for different
values of the resonance mass for flavor universal couplings 
(left plot), and flavor non-universal couplings (right plot, 
only at 90\% C.L.).}
\end{center}
\end{figure}

In Fig.~\ref{fig:anti} and Fig.~\ref{fig:universal}, we update 
our analysis from Ref.~\cite{Ferrario:2009bz,phd}
with the new measurement of the asymmetry in \Eq{renewcdf}. 
The new measurement do not disfavour completely axigluons 
(and colorons), at 95\% C.L.; there is still some room 
for negative (or vanishing) contributions beyond the SM (Fig.~\ref{fig:anti}). 
In the flavour universal scenario, Fig.~\ref{fig:universal} left,
large vector couplings at favoured at 90\% C.L., 
although at 95\% C.L. the allowed parameter space is much larger. 
In the flavour non-universal scenario, Fig.~\ref{fig:universal} right, 
there are not significant changes, although the allowed parameter 
space is again larger. We have not considered here possible further 
constrains from flavour observables~\cite{Chivukula:2010fk}.

\section{Coloured scalars}

Besides additional gauge bosons, Grand Unified Theories (GUT)
based on larger gauge groups, e.g., $SU(5)$, $SO(10)$, and $E6$, 
often introduce new coloured scalar states. For example, 
in $SU(5)$, the Higgs boson multiplets are made of the 
following field components
\bea
5_H &=& (1,2,1/2) + (3,1,-1/3)~, \nn \\
24_H &=& (8,1,0) + (1,3,0) + (3,2,-5/6) + (\bar{3},2,5/6) + (1,1,0)~, \nn \\
45_H &=& (8,2,1/2) + (\bar{6},1,-1/3) + (3,3,-1/3) + (\bar{3},2,-7/6)
+ (3,1,-1/3) \nn \\ && + (\bar{3},1,4/3) + (1,2,1/2)~.
\eea
Although, most of these states lie at, or close to, the unification 
scale, gauge coupling unification and proton decay might force some 
of them to be light~\cite{Perez:2008ry, Dorsner:2009mq}, 
and at reach at the LHC. 

Exchange of coloured scalars in the $s$-channel do not generate 
a charge asymmetry; hence t-channel contributions and thus 
flavour violating couplings need to be introduced 
to explain a large asymmetry.
Several authors~\cite{Dorsner:2009mq,Shu:2009xf,Arhrib:2009hu,Jung:2009pi,
Cao:2009uz} have considered scalar colour singlet $(1,2,-1/2)$, 
triplet $(\bar{3},1,4/3)$, sextet $(6,1,4/3)$ and octet $(8,2,-1/2)$ 
exchange in the $t$-channel. Other scalar states, like
$(6,3,1/3)$ and $(\bar{3},3,1)$, have not been analysed because
they are more constrained from flavour observables.
The most general up quark-top quark-scalar interaction 
is given by~\cite{Shu:2009xf}
\beq
{\cal L} = t^a \, 
\bar{t} (g_S + g_P \, \gamma_5) \, \phi^a \, u~.
\eeq
and the generated asymmetry depends only on the following 
combination of scalar $g_S$ and pseudoscalar $g_P$ 
couplings
\beq
y=\sqrt{g_S^2+g_P^2}~.
\eeq
In general, triplet~\cite{Dorsner:2009mq,Shu:2009xf,Arhrib:2009hu,
Cao:2009uz} and sextet~\cite{Shu:2009xf} appear to be in agreement 
with a large asymmetry, although requiring large flavour violating 
couplings, while singlet~\cite{Shu:2009xf,
Cao:2009uz} and octet~\cite{Dorsner:2009mq,Shu:2009xf} 
fail to accommodate the asymmetry. These models have to deal  
with potential $uu \to tt$, or same sign dileptons, which 
are quite constrained by Tevatron.

\section{Extra weak gauge bosons}

Extra weak gauge bosons appear in GUT, topcolor models, 
left-right models, or as Kaluza-Klein excitations of the 
SM weak bosons in extra dimensional models~\cite{Djouadi:2007eg}. 
The amplitude for top production through $Z'$ exchange in the
$s$-channel do not interfere with the SM amplitude
at hadron colliders, and thus its charge asymmetry is suppressed. 
In order to generate a large charge asymmetry several 
authors have considered $Z'$ and $W'$ in the 
$t$-channel~\cite{Cao:2009uz,Jung:2009jz,Cheung:2009ch,
Barger:2010mw,Cao:2010zb}. 
As for scalars, this requires to introduce large flavour 
violating couplings: 
\beq
{\cal L} = \bar{t} \, (g_V^{Z^{'}} + g_A^{Z^{'}}  \gamma_5) \, 
\gamma^\mu \, Z_\mu^{'} \, u
+ \bar{t} \, (g_V^{W^{'}} + g_A^{W^{'}}  \gamma_5) \, 
\gamma^\mu \, W_\mu^{'} \, d~.
\eeq
Furthermore, since weak bosons in the $t$-channel are more efficient 
than scalars in generating a large charge asymmetry, and in 
order to avoid $uu\to tt$ (same sign dileptons), the extra gauge
bosons need to be relatively light, having masses of the 
order of $200$~GeV.

\section{The charge asymmetry at the LHC}

Top quark production at the LHC is forward--backward symmetric in the 
laboratory frame as a consequence of the symmetric colliding
proton-proton initial state. 
Nevertheless, it is still possible to find a charge asymmetry
in suitable defined kinematic regions. 
QCD predicts that top quarks are preferentially emitted in 
the direction of the incoming quarks. But since quarks in 
the proton carry, on average, more momenta than antiquarks
the partonic asymmetry will be translated into an excess 
of top quarks in the forward and backward regions due 
to the boost into the laboratory frame~\cite{mynlo}. 
Similar arguments apply to the charge asymmetry 
generated at the partonic level from any other model. 
Thus, we define the integrated central charge
by selecting events in a given range of rapidity in 
the central region~\cite{Antunano:2007da,Ferrario:2008wm}:
\beq
A_C(y_C) = \frac{N_t(|y|\le y_C)-N_{\bar{t}}(|y|\le y_C)}
{N_t(|y|\le y_C)+N_{\bar{t}}(|y|\le y_C)}~.\label{eq:acyc}
\label{eq:central}
\eeq
The central asymmetry $A_C(y_C)$ obviously vanishes if the
whole rapidity spectrum is integrated, while a non-vanishing
asymmetry can be obtained over a finite interval of rapidity.

In contrast with Tevatron, top quark production at LHC is 
dominated by gluon-gluon fusion (70\% at 7~TeV and 90\% at 14~TeV), 
which is charge symmetric under higher order corrections.
The charge antisymmetric contributions to top quark
production are thus screened at LHC
due to the prevalence of gluon-gluon fusion.
This is the main handicap for that measurement.
The amount of events initiated by gluon-gluon collisions can nevertheless
be suppressed with respect to the $q\bar q$ and $gq(\bar q)$ processes,
the source of the charge asymmetry, by introducing a lower cut
on the invariant mass of the top-antitop quark system $m_{t\bar t}$;
this eliminates the region of lower longitudinal momentum
fraction of the colliding partons,
where the gluon density is much larger than the quark densities.
The charge asymmetry of the selected data samples is then enhanced,
although at the price of lowering the statistics. 

In Ref.~\cite{Ferrario:2008wm,phd} we have analyzed the magnitude 
of the asymmetry and its statistical significance at the LHC, in QCD and 
in the presence of a colour-octet vector boson (see \Eq{eq:G}). 
The statistical significance of the measurement can be 
maximised by tuning the maximum rapidity $y_C$ in \Eq{eq:central}
and by selecting events with a minimal top-antitop quark pair 
invariant mass, $m_{t\bar t}^{\rm min}$. We found that around
$y_C=0.7$ the statistical significance is maximised. In QCD, 
statistics compensate for the smallness of the charge asymmetry, 
and indeed it is not necessary to introduce any cut in 
$m_{t\bar t}$. In models with extra massive gluons, a cut 
at about half (or even below) of the mass of the heavy gluon 
that is probed maximises the statistical significance.
This is a very interesting feature because softer top and antitop 
quarks should be identified more easily than the very highly boosted 
ones~\cite{boosted}. 

The production of top quark pairs together with one jet
reach roughly half of the total inclusive cross-section calculated
at next-to-leading order (NLO)~\cite{Dittmaier:2008uj}.
The charge asymmetry in $t\bar t+$jet is produced by the interference
of initial- with final-state real gluon emission
(Fig. \ref{fig:aFBasym}). This charge asymmetry is of similar size,
but of opposite sign to the total $t\bar t$ inclusive asymmetry~\cite{mynlo}.
The exclusive charge asymmetry suffers, however, 
from huge higher order corrections~\cite{ttjetnlo}.
In Ref.~\cite{Ferrario:2009ee} we have extended our analysis 
to $t\bar t+$jet, particularly for Kaluza-Klein gluons 
where $g_A^q=0$ for light quarks~\cite{Randall:1999ee}, 
where the inclusive asymmetry vanishes at LO.
It is interesting to stress that,
contrary to the SM, where top quarks contribute to the asymmetry
only when they are in a colour-singlet state (colour factor equal
to $d_{abc}^2$), there are also colour-octet contributions proportional
to the colour factor $f_{abc}^2$ in $t\bar t+$jet.

\section{Summary}

The new measurements of the top quark charge asymmetry at the 
Tevatron reduce the discrepancy with the SM from 2$\sigma$
to 1.7$\sigma$, and do not disfavour completely vanishing or 
negative contributions beyond the SM at 95\% C.L. 
The new measurement thus relax some of the exclusion constrains 
obtained by several studies. 
We have updated our analysis for colour octet vector resonances   
and found that large vector coupling are still 
favoured at 90\% C.L. in flavour universal scenarios, with
a larger than before allowed parameter space at 95\% C.L. 
In flavour non universal scenarios, with $g_A^q g_A^t <0$
there is not a significant change, although again the 
allowed parameter space is slightly larger. 

From the analysis of other authors, scalar colour triplet and 
sextet states with large flavour violating couplings are
compatible with a large charge asymmetry, while colour singlet 
and octet fail to account for the data. Extra weak bosons 
in the $t$-channel again require large flavour violating 
couplings, and would exhibit masses close to the electroweak 
scale. Since the new measurement of the charge asymmetry is 
closer to the SM prediction, one can anticipate that smaller
flavour violating couplings will be needed to account for the
new measurement in these models. 

The measurement of the charge asymmetry from $t\bar t$
events, with or without associated jets, at the LHC seems 
promising, although challenging. The measurement requires to 
select relatively low boosted top quark events, 
which is certainly an advantage.
Although 1~fb$^{-1}$ should be enough for a first measurement, 
most probably several tens of fb$^{-1}$ of data will be 
necessary to distinguish among models.

\acknowledgments
This work is supported
by the Ministerio de Ciencia e Innovaci\'on under Grant
No. FPA2007-60323, by CPAN (Grant No. CSD2007-00042),
by the Generalitat Valenciana under Grant No. PROMETEO/2008/069,
and by the European Commission MRTN FLAVIAnet under Contract
No. MRTN-CT-2006-035482.

\end{document}